\documentstyle[pre,aps,multicol,epsfig]{revtex}

\begin{document}
\draft

\title{Correlation--Hole Method for Spectra of Superconducting 
       Microwave Billiards}

\author{H. Alt$^{1}$, H.-D. Gr\"af$^{1}$, 
        T. Guhr$^{2}$, H.L. Harney$^{2}$,\\
        R. Hofferbert$^{1}$, H. Rehfeld$^{1}$, 
        A. Richter$^{1}$ and P. Schardt$^{1,}$\footnote{
        Present address: Siemens AG, Bereich Medizinische Technik,
        D--91052 Erlangen, Germany}
       }
\address{$^{1}$
         Institut f\"ur Kernphysik, Technische Hochschule Darmstadt,\\
         D--64289 Darmstadt, Germany\\
         $^{2}$    
         Max--Planck--Institut f\"ur Kernphysik,\\
         D--69029 Heidelberg, Germany\\
        }

\date{\today}

\maketitle

\begin{abstract}
The spectral fluctuation properties of various two-- and
three--dimensional superconducting billiard systems are investigated by
employing the correlation--hole method. It rests on the sensitivity of
the spectral Fourier transform to long range correlations and is thus
an alternative technique to study chaotic dynamics.  First, we apply
the method to the eigenfrequencies which are extracted from the
measured resonances.  Second, we analyze the unfolded raw spectra,
including the shape of the resonances.  The merit of the method lies
in a clear separation of the statistics due to the positions and due
to the shape of the resonances. However, we show that statistical 
fluctuations of the intensities of the resonances have a strong impact
on the observable. Therefore, the visibility of the correlation hole
is studied as a function of the number of independent statistical
variables entering into the intensities. The visibility improves
if independent spectra are superimposed.
\end{abstract} 

\begin{multicols}{2}

\narrowtext

\section{Introduction}
\label{sec1}

In recent years, the experimental study of chaotic dynamics in
billiard systems has attracted considerable interest. Unlike nuclei,
atoms, molecules or solid state probes, billiard systems can be
specifically designed to investigate certain aspects of chaotic
dynamics. This ``toy model'' feature which the aforementioned systems
lack, makes billiard systems very useful for studies of chaotic
dynamics. Electromagnetic billiards simulating the corresponding
quantum systems were experimentally investigated for the first time in
Refs.~\cite{SS9092,DSF90,Sridhar91}.  The use of superconducting
instead of normally conducting billiards yields an immense improvement
in the quality of the measured spectra. First results were presented
in Ref.~\cite{Grapa}.

The fluctuation properties of a rich variety of systems in nuclear,
atomic, molecular and solid state physics have been studied
experimentally and theoretically. In the case of fully developed
chaos, they are found to be universal and very accurately described by
Random Matrix Theory~\cite{Porter,Mehta,BG84,Bohigas,Berry,Haake}.
Due to general symmetry constraints, a time--reversal invariant system
with conserved or broken rotational invariance is modeled by the
Gaussian Orthogonal (GOE) or the Gaussian Symplectic Ensemble (GSE),
while the Gaussian Unitary Ensemble (GUE) describes time--reversal
non--invariant systems~\cite{Mehta}.  Regular systems, on the other
hand, show significantly different fluctuation properties. Remarkably,
they are, unlike the chaotic ones, not generic and can differ from
system to system. Although one often encounters the complete lack of
any correlations, which is referred to as Poisson regularity, the
extreme opposite, i.e.~the totally correlated spectrum of the harmonic
oscillator, also falls into the regular class.

In order to study the fluctuation properties on short and long scales,
one commonly analyzes the nearest neighbor spacing distribution and
the spectral rigidity~\cite{BG84,Haake}. This requires to extract the
positions of the levels from the measured spectrum. Thus, the analysis
is done on the so--called ``stick--spectrum'', i.e. the sequence of
those eigenenergies or eigenfrequencies which could be identified in
the raw spectra. However, one often has to deal with poorly resolved
spectra which prevents some levels from being found. This
``missing level effect'' has a considerable impact since it
counterfeits correlations which do not exist.  Fluctuation measures
that are less sensitive to the missing level effect are therefore
highly desirable.

In molecular physics, such a technique was developed by Leviandier et
al.~\cite{Leviandier} for the analysis of the long range correlations.
The properly smoothed Fourier transform of the spectral
autocorrelation function maps the long range correlations onto short
scales in Fourier space.  As compared to fluctuations of regular
systems, chaotic dynamics causes a considerable suppression of this
Fourier transform near the origin, a so--called ``correlation hole''.
This has been experimentally observed in spectra of the molecules
acetylene, methylglyoxal and nitrogendioxyd~\cite{Leviandier,Delon}.
Recently, by re--analyzing the Nuclear Data Ensemble, it was shown that
nuclear spectra exhibit the correlation hole,
too~\cite{Leviandier,BohigasGSE}.

As pointed out already, fluctuation measures like the nearest neighbor
spacing distribution or the spectral rigidity can only be used for the
extracted stick--spectrum. The correlation--hole method of
Ref.~\cite{Leviandier}, however, is also applicable to the raw spectra
since the Fourier transform separates the statistics of the positions
of the eigenenergies from the statistics of the intensities and
widths.  A possibly faulty and incomplete extraction of the positions
and the widths from the spectra can thus be avoided.  Consequently,
the correlation--hole method is, in principle, less sensitive to the
missing level effect and it is worthwhile to study its applicability
to other physical systems.

The theory of the correlation hole for realistic spectra, i.e.
including the line widths, was worked out in Ref.~\cite{Guhr1} in the
framework of a scattering model. In the same picture, the correlation
hole observed by laser--spectroscopy in methylglyoxal~\cite{Leviandier}
was numerically simulated in Ref.~\cite{Guhr2}. A summary and a
qualitative discussion of the Fourier transform of statistical spectra
can be found in Ref.~\cite{LPLBS}. A conspicuous short version of the
theory of the correlation hole is presented in Appendix A of
Ref.~\cite{LoSe}.  In Ref.~\cite{Alhassid}, the correlation hole is
related to its classical analogue, the survival probability. The
correlation--hole method was first applied to billiard spectra by
Kudrolli et al.~\cite{Sridhar}.  However, these authors studied only
the extracted stick--spectra.

In the present work, we apply this method to the rich variety of
billiard spectra that have been measured in Darmstadt over the last
years. We have two goals: First, we want to verify the significance of
the correlation hole for billiards of quite different geometries by
analyzing the stick--spectra of the extracted levels.  Second, we use
the correlation--hole technique for the unfolded raw spectra and
discuss the merits and the problems of such an analysis.  We present a
method of superimposing several raw spectra in order to improve upon
the visibility of the correlation hole.  This is the first time that such
a detailed study is performed for billiard systems.  The excellent
resolution of the spectra measured in superconducting microwave
cavities makes the Darmstadt data the ideal object for such an
analysis.

After a short description of the experiment in Sec.~\ref{sec2}, we
present the theoretical concepts in Sec.~\ref{sec3}. We perform the
analysis of our data in Sec.~\ref{sec4} and finish with conclusions in
Sec.~\ref{sec5}.

\section{Experiment} 
\label{sec2}

Due to the equivalence of the stationary Schr\"odinger equation for
quantum systems to the corresponding Helmholtz equation for
electromagnetic resonators in two dimensions, it is possible to
simulate a quantum billiard of a given shape with the help of a
sufficiently flat macroscopic electromagnetic resonator of the same
shape~\cite{SS9092,DSF90,Sridhar91}.

\begin{figure}  
\centerline{\epsfxsize=8.6cm
\epsfbox{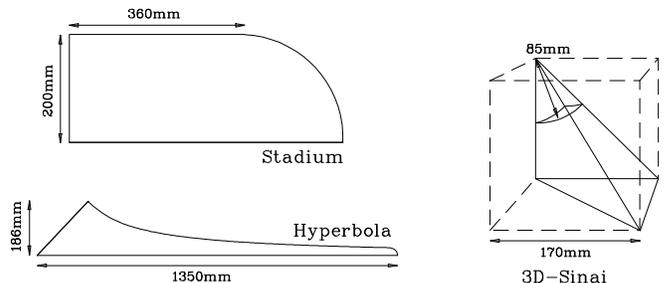}}
\caption{
The geometries of the superconducting resonators which
were used in the experiments. Note that all systems are desymmetrized
to avoid parity--mixing. In comparison with the full systems this
yields a quarter stadium, an eighth of a hyperbola and a 48--th of a
3D--Sinai--billiard, respectively. In the case of the latter, the sides
of the drawn cube are 170~mm long and the radius of the removed sphere
measures half of this length, i.e.~85~mm.
}
\protect\label{fig1}
\end{figure}

Since 1991, we have experimentally studied several two-- and
three--dimensional billiard systems using superconducting microwave
resonators made of niobium.  Figure~\ref{fig1} shows some of the
investigated cavities as well as their dimensions.  First experiments
using superconducting instead of normally conducting resonators were
performed in a desymmetrized Bunimovich--stadium--billiard and a
truncated hyperbola--billiard~\cite{Grapa,Hyperbel,widths} using the
2K--cryostats of the superconducting electron linear accelerator
S--DALINAC~\cite{Beschl} in Darmstadt.  Recently, a desymmetrized
3D--Sinai--billiard~\cite{Primack,3DSinai} was experimentally
investigated in a new and very stable 4K--bath--cryostat.  Note that
the electromagnetic Helmholtz equation is vectorial in three
dimensions and cannot be reduced to an effective scalar equation.
Thus, it is structurally different from the scalar Schr\"odinger
equation.  It is of considerable interest that the statistical
concepts developed in the theory of quantum chaos and random matrices
are also applicable to arbitrary classical electromagnetic wave
phenomena. Conceptually, this is similar to the study of spectral
fluctuations of elastomechanical eigenmodes in
aluminum~\cite{Weaver,ACME} and quartz~\cite{quartz} blocks, which are
also described very well by Random Matrix Theory.

\begin {figure}
\centerline{\epsfxsize=8.6cm
\epsfbox{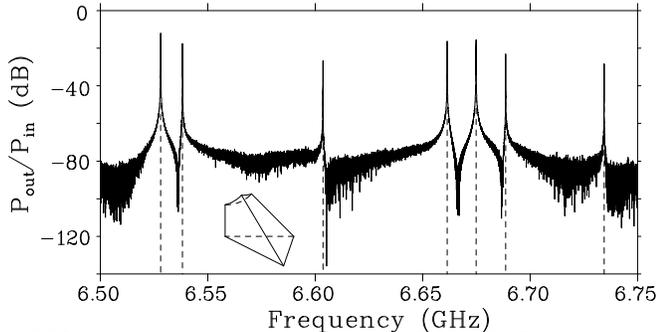}}
\caption{
Transmission spectrum of the 3D--Sinai--billiard in the range between
6.50 and 6.75~GHz. The signal is given as the ratio of output power to
input power on a logarithmic scale. The dashed vertical lines mark the
extracted eigenfrequencies $f_\mu$ as used in the stick--spectrum.
}
\label{fig2}
\end{figure}

All resonators mentioned above were excited in the frequency range
between 0 and 20~GHz using capacitively coupling dipole antennas
sitting in small holes on the niobium surface. Using one antenna for
the excitation of the resonator and either another or the same one for
the detection of the microwave signal, we were able to measure the
transmission or the reflection spectrum of the resonator,
respectively, by employing a Hewlett Packard HP8510B vector network
analyzer. As an example, Fig.~\ref{fig2} shows a typical transmission
spectrum of the 3D--Sinai--billiard in the range between 6.50 and
6.75~GHz.  The signal is given as the ratio of output power to input
power on a logarithmic scale.  The measured resonances have quality
factors of up to $Q\approx 10^7$ and signal--to--noise ratios of up to
$S/N\approx 70~{\rm dB}$ which makes it easy to separate the
resonances from each other and from the background.  As a consequence,
all the important characteristics like eigenfrequencies and widths can
be extracted with a very high
accuracy~\cite{widths,Brentano,Brentanoneu}.  A detailed analysis of
the original spectra yields a total number of approximately 1000
resonances for the 2D--billiards, hyperbola and stadium, and nearly
1900 resonances for the 3D--Sinai--billiard. These eigenvalue sequences
form the basis of the present test of the correlation--hole method.

\section{Theoretical Concepts of the Correlation--Hole Method} 
\label{sec3}

We summarize, for the convenience of the reader, the basic ideas in
Sec.~\ref{sec3a} and some earlier results on the two--level form
factors in Sec.~\ref{sec3b}. We discuss the theoretical results on the
correlation hole for stick-- and raw spectra in Secs.~\ref{sec3c} and
\ref{sec3d}, respectively.

\subsection{Basic Ideas}
\label{sec3a}

Consider a spectrum $I(E)$ or $I(f)$ measured as a function of energy
$E$ or, in our case, frequency $f$.  This spectrum is described as a
finite superposition of isolated or interfering resonances of a given
universal shape $L(f)$ with statistically distributed positions,
intensities and widths.  Although the shape is often known to be a
Lorentzian~\cite{Brentano,Brentanoneu}, we keep the discussion
general. In order to analyze the true fluctuations, one removes
secular variations of the level density, i.e. the Weyl-- or
Thomas--Fermi contribution~\cite{Weyl1,Weyl2,BalHilf}.  To this end,
one introduces~\cite{BG84} the smooth part $N^{\rm Weyl}(f)$ of the
integrated level density as the new coordinate by setting $x=N^{\rm
  Weyl}(f)$ for our experimental and theoretical discussions.  Since
we wish to study generic fluctuations, we shall henceforth assume that
this unfolding procedure has been performed and write $I=I(x)$. This
implies that, in the variable $x$, the mean level spacing is unity
everywhere.

Hence, we study a spectrum of $N$ levels $x_\mu,\ \mu=1,\ldots,N$
on this unfolded scale. In the following, the level number $N$ is always
assumed to be large. To simplify the theoretical description, 
we make the further assumption that the line shape $L(x)$ does not
depend on the intensities $y_\mu$ for a given resonance $\mu$
in the expression
\begin{equation}
       I(x)=\sum_{\mu=1}^N y_\mu \, L(x-x_\mu) \ ,
       \label{intensity}
\end{equation}
where the line shape is normalized to unity.
The observable of interest is the decay function, i.e.~the modulus
squared of the Fourier transform of the spectrum
\begin{equation}
      |C(t)|^2 = \Bigg|\int\limits_{-\infty}^{+\infty} dx \,
                 I(x) \, \exp(2\pi i x t)\Bigg|^2 \ .
\label{autocorri}
\end{equation} 
The Fourier coordinate $t$ defines an unfolded time.
This expression can be rewritten as the Fourier transform
\begin{equation}
      |C(t)|^2 = \int\limits_{-\infty}^{+\infty} 
                   d\omega \, A(\omega) \, \exp(2\pi i \omega t)
\label{autofour}
\end{equation}
of the autocorrelation function
\begin{equation}
      A(\omega) = \int\limits_{-\infty}^{+\infty} d\Omega \, 
                        I(\Omega-\omega/2) \, I(\Omega+\omega/2) \ .
\label{auto}
\end{equation}
At this point a comment on the assumed independence of the the line
shape $L(x)$ of the intensities $y_\mu$ is in order.  From scattering
theory~\cite{MW} it is known that this is the case only if the number
$\Lambda$ of open decay channels is very large. Strictly speaking,
this is not true for the experiments to be analyzed here. The decay
channels are the antennas that couple the cavity to the external world
and, hence, $\Lambda= 2-4$, see the discussion below.  However, the
assumption of independence is to some extent justified here since it
affects only the long time behavior of the decay function $|C(t)|^2$,
whereas the correlation hole is a feature found in the short time
behavior. Moreover, the more general theory does also show that the
neglect of interference effects as done in Eq.~(\ref{intensity}) is
justified.  Thus, it turns out that our simplification does still
yield a satisfactory description of the correlation hole in all cases
we studied.

Replacing the energy average in Eq.~(\ref{auto}) by the average over
an ensemble of resonances, one can make use of two results of Random
Matrix Theory~\cite{Porter,Mehta,widths}: (i) the $y_\mu$ are
independent of the $x_\mu$ and (ii) for $\mu\ne\nu$ the intensities
$y_\mu,y_\nu$ are independent of each other. The function~(\ref{auto})
then can be cast into the form
\begin{eqnarray}
A(\omega) & = & N\overline{y^2}
                \int\limits_{-\infty}^{+\infty} 
                dx' \, L(\Omega-\omega/2-x') \,
                L(\Omega+\omega/2-x') \nonumber\\
          & & + \, N\overline{y}^2 \nonumber\\
          & & - \, N\overline{y}^2 
                \int\limits_{-\infty}^{+\infty} dx' \,
                \int\limits_{-\infty}^{+\infty} dx'' \,
                L(\Omega-\omega/2-x') \nonumber\\
          & & \qquad \times \, L(\Omega+\omega/2-x'') \, Y_2(x''-x') \ .
\label{autoresult} 
\end{eqnarray} 
Here, $Y_2(x)$ denotes the Dyson--Mehta two--level cluster
function~\cite{Mehta}. It describes the two--point correlations, more
precisely, $(1-Y_2(x))dx$ is the probability to find two resonances
separated by the distance $x$ on the unfolded scale. Thus, for large
arguments $x$, the correlations have to disappear and $Y_2(x)$
approaches zero.  Note that $\overline{y}^2$, the square of the first
moment of the intensities, and $\overline{y^2}$, the second moment
appear in Eq.~(\ref{autoresult}).

We introduce the Fourier transform of the two--level cluster function,
\begin{equation}
         b_2(t) = \int\limits_{-\infty}^{+\infty} dx \,
                          Y_2(x) \, \exp(2\pi i x t) \ ,
\label{Y2four}
\end{equation}
and analogously the Fourier transform $\tilde{L}(t)$ of the line shape
$L(x)$. The function $b_2(t)$ is referred to as the two--level form
factor~\cite{Mehta}.  Just like the two--level cluster function, it has
to vanish for large arguments $t$.  Since the last term on the right
hand side of Eq.~(\ref{autoresult}) contains a convolution, we can
make use of the convolution theorem to evaluate the Fourier transform
of $A(\omega)$. Collecting everything, one
finds~\cite{Leviandier,Guhr1,LPLBS,LoSe} for non--negative times
\begin{equation}
       |C(t)|^2 = N\overline{y^2} \delta(t) 
                  \, + \, N\overline{y^2} \, |\tilde L(t)|^2 \,
                          \left(1-\alpha b_2(t)\right) \ ,
\label{rawfour}
\end{equation}
where we denote the ratio of the statistical moments by
\begin{equation}
     \alpha = \overline{y}^2 \, / \, \overline{y^2} \ .
\label{alpha}
\end{equation}
The correlation hole is described by the function $\big(1-\alpha
b_2(t)\big)$, while $|\tilde L(t)|^2$ describes the decay of the
resonances.  Thus, the statistics of the positions is separated from
the statistics of the intensities. The $\delta$--function in
Eq.~(\ref{rawfour}) occurs since we assumed $N$ to be very large.  For
a finite number of levels, this contribution will acquire a width as
discussed in detail in Refs.~\cite{Guhr1,LoSe}.  Due to the high
number of levels in our data, we disregard this contribution in the
sequel.

If the resonances are well isolated, then $|\tilde L(t)|^2$ varies
much more slowly than $b_2(t)$. In the extreme case of vanishing
widths, i.e. $L(x) = \delta(x)$ the function~(\ref{rawfour})
approaches $N\overline{y^2}$ for large times $t$.  For the more
realistic Lorentzian line shape~\cite{Brentano,Brentanoneu} one has
$\tilde{L(t)} = \exp(-\pi\Gamma t)$, where $\Gamma$ is the total
width, implying that the function~(\ref{rawfour}) decays
exponentially.

\subsection{Correlation Hole and Two--Level Form Factors}
\label{sec3b}

For the convenience of the reader, we collect here the well known
results for the form factors $b_2(t)$ introduced in
Eq.~(\ref{Y2four}).

The case that the positions $x_\mu,x_\nu$ of any two different
resonances $\mu \neq \nu$ are completely uncorrelated is referred to
as Poisson regularity~\cite{Mehta,BG84}. Obviously, the two--level
cluster function must be zero everywhere, $Y_2^{\rm Poisson}(x) = 0$,
and therefore also $b_2^{\rm Poisson}(t)=0$ which results in $|C(t)|^2
\approx N \overline{y^2} = {\rm const.}$ even for small values of $t$.
There is no correlation hole.

\begin {figure}
\centerline{\epsfxsize=8.6cm
\epsfbox{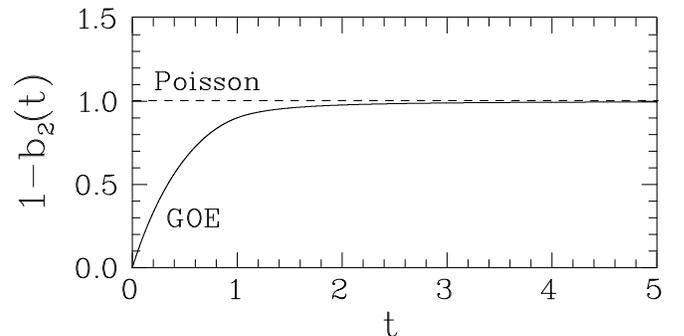}}
\caption{
The phenomenon of the correlation hole due to the two-level form
factor is clearly visible for small values of $t$.  The function
$1-b_2(t)$ is displayed for the cases of Poisson and GOE statistics
according to Eq.~(\protect\ref{b2GOE}).
}
\label{fig3}
\end{figure}

In the case of fully developed chaos, the statistics of the positions
is described by the Gaussian Ensembles~\cite{Mehta}.  The general
symmetry constraints imply that a time--reversal invariant system with
conserved or broken rotational invariance is modeled by the Gaussian
Orthogonal or the Symplectic Ensemble, while the Gaussian
Unitary Ensemble describes time--reversal non--invariant
systems. We summarize the results for the form factors~\cite{Mehta}.
The situation most commonly encountered is described by the
GOE, yielding
\begin{eqnarray}
& & b_2^{\rm GOE}(t) = \nonumber\\
& & \qquad \ \cases{ 1-2t+t\ln(1+2t), & $0 < t \le 1$ \cr
                   -1+t\ln((2t+1)/(2t-1)), & $t > 1$ } \ .    
\label{b2GOE}
\end{eqnarray}
This function is displayed in Fig.~\ref{fig3}. It corresponds to a
decay function~(\ref{rawfour}) for $\alpha=1$ which has been
normalized by $N\overline{y^2}$.  The GSE form factor is given by
\begin{eqnarray}
& & b_2^{\rm GSE}(t) = \nonumber\\
& & \qquad\qquad \cases{ 1-t/4+(t\ln|1-t|)/4, & $0< t \le 2$ \cr
                   0, & $t > 2$ } \ ,
\label{b2GSE}
\end{eqnarray}
and thus exhibits a divergence at $t=1$.  Importantly, for all fully
chaotic cases, we have for vanishing times $b_2(0)=1$. Thus, according
to Eq.~(\ref{rawfour}), the deepest point of the correlation hole is
reached for small values of the time $t$ and has the value $|C(t)|^2
\approx N\overline{y^2} (1-\alpha)$.

\subsection{Stick--Spectra}
\label{sec3c}

Suppose that all resonances have a vanishing width, i.e.~we consider
a sequence of levels at positions $x_\mu$
represented by $\delta$--functions. If, moreover, all of them
have the same intensity, $y_\mu = \overline{y}$,
Eq.~(\ref{intensity}) reduces to
\begin{equation}
  I(x) = \overline{y} \sum_{\mu=1}^N \delta(x-x_\mu)
\label{sticky}
\end{equation}
which is nothing but the spectral function.  It serves as the
mathematical definition of the term ``stick--spectrum''~\cite {Delon}.
Formally, the probability distribution of the intensity $y_\mu$ is
given by a $\delta$--distribution which implies
$\overline{y^2}=\overline{y}^2$, hence, according to the
definition~(\ref{alpha}), we find $\alpha=1$.  Consequently, the
function $|C(t)|^2$ of Eq.~(\ref{rawfour}) becomes
\begin{equation}
|C(t)|^2 = N\overline{y}^2 \, 
          \left( \delta(t)+1-b_2(t) \right) \ .
\label{average}
\end{equation}
This is the case of maximum visibility of the correlation hole.  For a
realistic distribution $p(y)$ of the intensities, we will always find
values of $\alpha$ which are smaller, often considerably smaller, than
unity. 

\subsection{Correlation Hole for Raw Spectra} 
\label{sec3d}

The typical raw spectra of our resonators can be described as a superposition 
of isolated Lorentzian resonances \cite{Brentano,Brentanoneu}
with statistically distributed locations, intensities and widths.
This closely parallels the situation of isolated resonances
in a compound nucleus scattering experiment~\cite{MW}.
Due to a minimized surface resistance in the superconducting billiards,
the total width is basically composed of a sum over partial widths which
describe the power dissipation into the $\Lambda$ decay channels realized
by the antennas. Thus, each total width is given by
\begin{equation}
      \Gamma_\mu=\sum_{c=1}^\Lambda \Gamma_{\mu c}
      \label{totalwidth}
\end{equation}
for every resonance labeled by $\mu$ and channels denoted by $c$.
The small number of decay channels $\Lambda$ in our experiments, i.e
$\Lambda=2-4$, leads to a set of strongly fluctuating widths
$\Gamma_\mu$.  We remark that this allows to examine quantum phenomena
like the non--exponential decay of the spectral autocorrelation
function~\cite{widths}.

The simplified theoretical model which was introduced in
Sec.~\ref{sec3a}, however, is by construction not suited to describe
this general case but rather explains spectra of systems with a
constant total width. This situation can arise due to a high number of
decay channels~\cite{MW} or due to certain peculiarities of the
physical system or the measurement~\cite{Leviandier,Guhr2}.  As
mentioned above, the present simplification concerns the long time
behavior of $|C(t)|^2$ and not the correlation hole which shows up on
short time scales. This simplification is not valid if there are
sizeable correlations between the intensities $y_\mu$ and the total
widths $\Gamma_\mu$. Therefore, in the present discussion, it is
sufficient to assume a realistic distribution for the intensities
$y_\mu$.  This defines an adequate model for raw spectra and has to be
viewed in contrast to the $\delta$--distribution of the intensities
$y_\mu$ considered in Sec.~\ref{sec3c}.  It should be emphasized that
we, for reasons of consistency, neglect interference terms between the
individual resonances. Hence, in the terminology of scattering theory,
our model applies to the case of isolated resonances.

If the spectrum $I(x)$ has been measured in reflection in, say, channel $a$,
the intensities $y_\mu$ of the $\mu$--th resonance are given by
\begin{mathletters}
\label{dreizwanz}
         \begin{equation}
                y_\mu=\Gamma_{\mu a} \ ,
         \label{reflint}
         \end{equation}
where $\Gamma_{\mu a}$ is the partial width of the resonance with
respect to the channel, i.e.~the antenna, $a$.  Experiments as well as
the theory of random matrices show that the partial widths are
distributed according to a Porter--Thomas 
law~\cite{Porter,Mehta,widths}, which describes a fully chaotic
system, resulting in
         \begin{equation}
                \alpha = 1 \, / \, 3 \ .
         \label{reflalpha}
         \end{equation}
\end{mathletters}
If, however, the spectrum $I(x)$ has been measured in transmission
from channel $a$ to, say, channel $b$, the intensities $y_\mu$ are
given by the products
\begin{mathletters}
\label{vierzwanz}
         \begin{equation}
                y_\mu = \Gamma_{\mu a} \, \Gamma_{\mu b}
         \label{tranint}
         \end{equation} 
of the partial widths with respect to the entrance and exit channels.
Taking $\Gamma_{\mu a}$ and $\Gamma_{\mu b}$ to be statistically
independent variables with Porter--Thomas distributions one arrives at
         \begin{equation}
                \alpha = 1 \, / \, 9 \ .
         \label{tranalpha}
         \end{equation}
\end{mathletters}
These two values of $\alpha$ simply reflect a Gaussian distribution
for the decay amplitudes which is at the core of the Porter--Thomas law.

Hence, as indicated already, the statistical fluctuations of the
weight $y$ more or less suppress the correlation hole. The
$\delta$--distribution considered in Sec.~\ref{sec3c} is much more
favorable for the visibility of the correlation hole than a realistic
distribution which will always give $\alpha < 1$.  Reducing these
fluctuations would restore the correlation hole. This can be achieved
by the superposition of statistically independent spectra. Suppose
that spectra have been measured via all, or all possible, combinations
of the $\Lambda$ antennas attached to a given resonator. The positions
$x_\mu$ of the resonances are the same in all spectra, the intensities
$y_\mu$, however, vary from spectrum to spectrum. Thus, the
intensities of the spectrum obtained from superimposing all these
spectra will fluctuate much less. In the limit of a superposition of
infinitely many spectra, all intensities will be the same and we are
back to the case of the stick--spectrum.

We want to make this argument more quantitative. First, we discuss
reflection measurements. Write the $\mu$--th intensity of the
superposition of all possible reflection spectra in the form
\begin{mathletters}
    \label{fuenfzwanz}
    \begin{equation}
         y_\mu = \sum_{a=1}^\Lambda\Gamma_{\mu a} \ .
    \label{superreflint}
    \end{equation}
Under the assumptions that
$\Gamma_{\mu a}$ has Porter--Thomas statistics and that the
average value is independent of $\mu$ we find
    \begin{equation}
       \alpha = \frac{\Lambda}{\Lambda+2} \ .
       \label{reflalphatot}
    \end{equation}
\end{mathletters}
For $\Lambda=1$ one recovers the result~(\ref{reflalpha})
and, as expected, this expression approaches unity for large $\Lambda$.

Second, we turn to transmission measurements. Adding up all possible 
transmission spectra, we obtain the $\mu$--th intensity
\begin{mathletters}
    \label{sechszwanz}
    \begin{equation}
         y_\mu = \frac{1}{2}\sum_{{a,b=1} \atop {a\ne b}}^\Lambda
         \Gamma_{\mu a} \, \Gamma_{\mu b} \ .
    \label{supertranint}
    \end{equation}
Under the same assumptions as above this leads to
      \begin{equation}
      \alpha =  \frac{\Lambda-1}{\Lambda+7} \ .
      \label{tranalphatot}
      \end{equation}
\end{mathletters}
Again, this is consistent with the result~(\ref{tranalpha})
for $\Lambda=2$. Moreover, for large $\Lambda$, 
the expression~(\ref{tranalphatot}) approaches unity, as it should be.

\section{Application to Experimental Data}
\label{sec4}

After some general considerations in Sec.~\ref{sec4a}, we present the
analysis of stick-- and raw spectra in Secs.~\ref{sec4b} and
\ref{sec4c}, respectively.

\subsection{General Considerations}
\label{sec4a}

The Fourier transform $C(t)$ of the measured spectrum $I(x)$ does
still contain the entire information. As is well known, if the
experimental data consist of many levels in a sufficiently long
interval, the Fourier transform can, before unfolding, be used to
obtain information about the periodic orbits of the
system~\cite{Gutzwiller} which manifest themselves in a rich structure
consisting of many peaks. Here, however, we aim at an understanding of
the generic statistical features of our experimental data. In other
words: since we are not interested in resolving individual properties
like periodic orbits, we have to average over all realizations of the
physical system in question.  This, however, is precisely what Random
Matrix Theory did for us when we went from the autocorrelation
function~(\ref{auto}) to its ensemble average. According to Delon et
al.~\cite{Delon} the ensemble average can be simulated by applying a
smoothing procedure to the experimental decay function $|C(t)|^2$.  It
turns out~\cite{Delon} that the most appropriate procedure is a
convolution of $|C(t)|^2$ with a Gaussian. Hence, we have to compare
the theoretical results to the function
\begin{equation}
  \left\langle|C(t)|^2\right\rangle = 
      \int\limits_{-\infty}^{+\infty} dt' \, 
           |C(t')|^2 \, \frac{1}{\sqrt{2\pi\sigma_t^2}}
          \exp\left(-\frac{(t-t')^2}{2\sigma_t^2}\right) \ ,
\label{fullGauss}
\end{equation}
where the variance $\sigma_t$ was chosen to depend on the time 
as $\sigma_t=t/10$.
This procedure is referred to as ``full Gaussian smoothing''.

\subsection{Correlation Hole for Stick--Spectra}
\label{sec4b}

We now analyze the sequences of levels which were extracted from the
measured spectra, i.e.~the stick--spectra.  

In some cases, we perform an additional test of GOE characteristics
following the discussion of Ref.~\cite{BohigasGSE}.  A theorem due to
Dyson and Mehta~\cite{DysonMehta} states: If one takes a GOE spectrum
$\{f_1,f_2,f_3,...\}$ and divides it into two sequences of odd and
even indices, i.e.~into $\{f_1,f_3,f_5,...\}$ and
$\{f_2,f_4,f_6,...\}$, then each of these two spectra obeys, after
proper unfolding, the GSE statistics. Following this idea, the
experimentally found stick--spectrum was divided into two equivalent
sets with half the number of eigenfrequencies, put together
sequentially and the whole analysis was repeated.  The theoretical
prediction is given by using Eq.~(\ref{b2GSE}) in the
expression~(\ref{average}).  In contrast to the GOE case, the 
GSE form factor $b_2(t)$ has a singularity at $t=1$ due to the
different oscillatory structure of the two--level correlation function
$Y_2(x)$. Thus, one expects a characteristic peak in the function
$\big<|C(t)|^2\big>$ at this time $t$ which gives information on
correlations on the scale of about two mean level spacings. 
As pointed out in Ref.~\cite{BohigasGSE}, the decay function
constructed in this way by omitting every other level contains
information on higher than two--level correlations of the original
spectrum.

In Secs.~\ref{sec4b1} and \ref{sec4b2}, we discuss the hyperbola and
the stadium--billiard, respectively. In Sec.~\ref{sec4b4}, we analyze
the stick--spectrum of the 3D--Sinai--billiard.

\subsubsection{Hyperbola--Billiard}
\label{sec4b1}

\begin {figure}
\centerline{\epsfxsize=8.6cm
\epsfbox{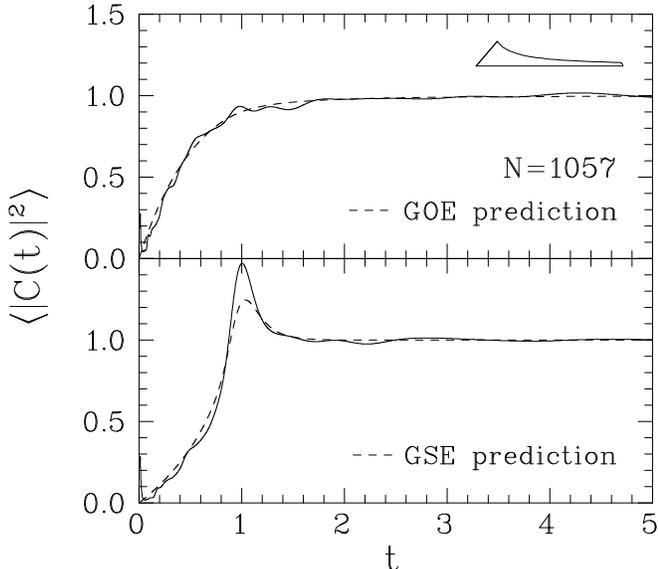}}
\caption{
The function $\langle |C(t)|^2 \rangle$ for the stick--spectrum of the
hyperbola-billiard. Note that the ordinate is divided by
$N\overline{y}^2$ in order to allow immediate comparison with Fig.~3
by using Eq.~(\protect\ref{average}). The full line is the
experimental and the dashed line the theoretical result.
}
\label{fig4}
\end{figure}

The upper part of Fig.~\ref{fig4} shows the result for the measured
spectrum of the hyperbola-billiard whose shape is displayed in
Fig.~\ref{fig1}.  The full line is the experimental result for
$\langle|C(t)|^2\rangle$ according to Eq.~(\ref{fullGauss}).

Since the hyperbola-billiard has been proven~\cite{Sieber,Hesse} to be
fully chaotic in the classical limit, one expects~\cite{BerryGOE} to
find GOE fluctuations in the spectrum. Indeed, as Fig.~\ref{fig4}
shows, the agreement with the theoretical prediction of
Eq.~(\ref{average}) with $b_2(t)$ given by Eq.~(\ref{b2GOE}) is very
good.  Note that $\big<|C(t)|^2\big>$ as well as Eq.~(\ref{average})
have been divided by $N\overline{y}^2$ to allow a comparison with
Fig.~\ref{fig3}.

The test of GSE statistics according to the Dyson-Mehta observation
mentioned above yields the curve shown in the lower part of
Fig.~\ref{fig4}.  Again, the agreement with the theory is good.  Here,
to be consistent, we have smoothed the theoretical curve, too, by help
of Eq.~(\ref{fullGauss}). We do not exclude the possibility that the
discrepancy between this curve and the experimental result at $t=1$
might hint at deviations of the higher order and the medium range
correlations in the spectrum from the GOE prediction. Due to the
limited amount of data, however, we cannot perform more detailed tests
which would be necessary to make a definite statement.

\subsubsection{Stadium--Billiard}
\label{sec4b2}

\begin {figure}
\centerline{\epsfxsize=8.6cm
\epsfbox{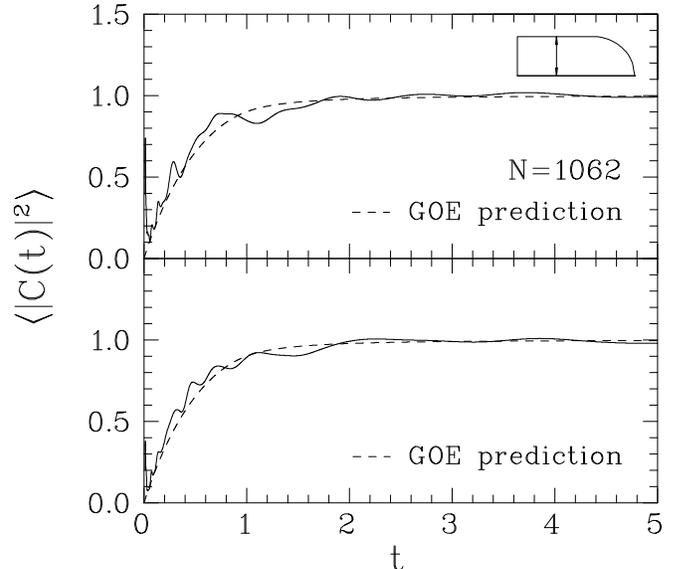}}
\caption{
The function $\langle |C(t)|^2 \rangle$ for the stick--spectrum of the
stadium-billiard, as in Fig.~4. The experimental curve (solid line) in
the upper part includes the bouncing ball contribution. In the lower
part, the bouncing ball contribution has been removed. The theoretical
prediction (dashed line) is the same in both cases.
}
\label{fig5}
\end{figure}

The Bunimovich-stadium-billiard~\cite{Bunimovich} displayed in
Fig.~\ref{fig1} is also totally chaotic in the classical sense but it
has an additional feature: In contrast to the hyperbola, the stadium
possesses one neutrally stable and non-isolated periodic orbit, the
so-called bouncing ball orbit which propagates between the two
straight and parallel segments of the geometry. This lends a certain
type of regular characteristics to the system. Thus, in calculating
$\langle|C(t)|^2\rangle$, we test the influence of this remaining
regularity on the correlation hole.  The result is given in Fig.~\ref{fig5},
where in the lower part the bouncing ball orbit has been removed by
extracting the term that the bouncing ball orbit contributes to the
smooth part of the level density~\cite{Smilansky} together with the
common Weyl contribution in the unfolding procedure.  As can be seen
from the figure, this extraction amounts to a very slight correction
towards the pure GOE characteristics. Due to the mapping of long range
spectral properties onto short times, the correlation hole is quite
insensitive to the bouncing ball orbit.

\subsubsection{3D--Sinai--Billiard}
\label{sec4b4}

\begin {figure}
\centerline{\epsfxsize=8.6cm
\epsfbox{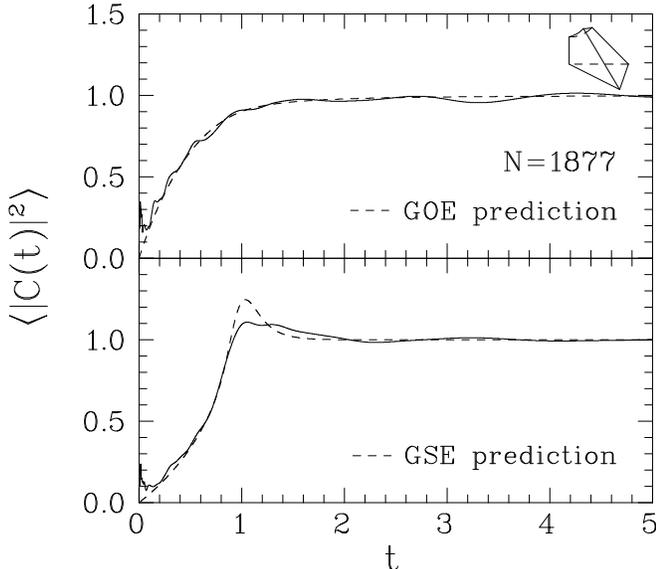}}
\caption{
The function $\langle |C(t)|^2 \rangle$ for the stick--spectrum
of the 3D-Sinai-billiard, as in Fig.~4.
}
\label{fig8}
\end{figure}

Finally, we have considered a spectroscopic system that generalizes
the statistical concepts of quantum chaos in the sense that, although
it does not represent or simulate a quantum system, its spectral
features are found to coincide with GOE characteristics. As in the
case of elastomechanical eigenmodes~\cite{Weaver,ACME,quartz}, this
indicates that many classical wave phenomena might follow the
predictions of Random Matrix Theory. The system at hand is the
3D-Sinai-billiard~\cite{Primack,3DSinai,ACME,quartz} shown in
Fig.~\ref{fig1}. Its wave dynamics is described in terms of the
vectorial Helmholtz equation in three dimensions.  The resulting
function $\langle |C(t)|^2 \rangle$ is displayed in Fig.~\ref{fig8}.
As in the case of the hyperbola-billiard, we also performed the GSE
test on our data.

Note that, in this geometry, the eigenvalues of the quantum mechanical
Schr\"odinger equation as well as the eigenvalues of the vectorial
electromagnetic Helmholtz equation show, due to the bouncing ball
orbits, slight deviations from pure GOE behavior~\cite{Primack} in
other statistics like the number variance $\Sigma^2$ or the spectral
rigidity $\Delta_3$. It is not possible to draw direct conclusions
from that to the present system since the ray limit of the vectorial
Helmholtz equation is different from the classical limit of the
Schr\"odinger equation.  
However, it could be speculated that the
influence of the corresponding bouncing ball orbits in the present
system is suppressed due to an effective average over all their
different lengths. In any case, the effect of the bouncing ball orbits
becomes visible at comparatively large lengths in the spectrum. Thus,
in our observable, it cannot be extracted with statistical
significance since the Fourier transform maps it onto values of $t$
which are of the order of the inverse length in the spectrum. Remember
that the correlation hole is found at values of $t$ which are roughly
of the order unity. Nevertheless, a suppression of the peak at $t=1$
in the experimental curve is seen in the GSE test in the lower part of
Fig.~\ref{fig8}. One might be tempted to interpret this as due to
those effects. Again, more detailed tests would require more data and,
in this particular case, a thorough theoretical discussion of the ray
limit of the vectorial Helmholtz equation, too.

\subsection{Raw Spectra}
\label{sec4c}

As far as the stick--spectra are concerned, the phenomenon of the
correlation hole is obviously well understood. The agreement with
theoretical predictions for all considered systems is satisfactory. We
now apply the method to original and idealized raw spectra in
Secs.~\ref{sec4c1} and \ref{sec4c2}, respectively.  In
Sec.~\ref{sec4c3}, the influence of certain statistical
fluctuations on our findings is discussed and demonstrated using
synthetic spectra.

\subsubsection{Original Raw Spectra}
\label{sec4c1}

\begin {figure}
\centerline{\epsfxsize=8.6cm
\epsfbox{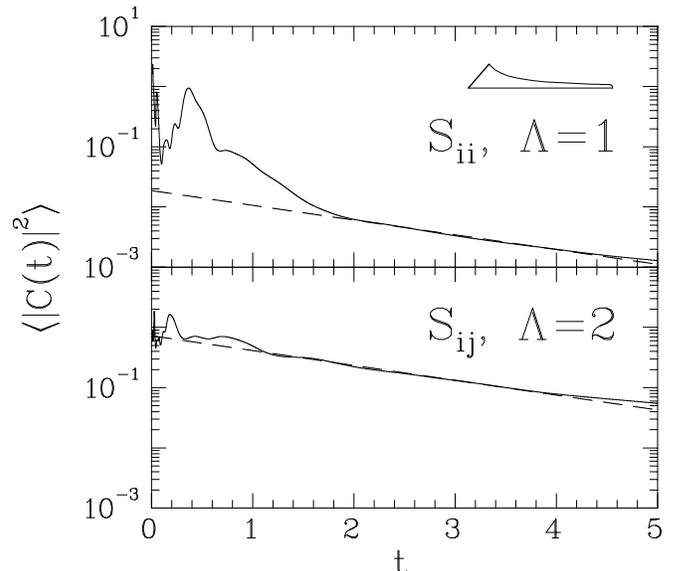}}
\caption{
The function $\langle |C(t)|^2 \rangle$ for original raw spectra of
the hyperbola-billiard.  The two parts of the figure show the result
from a spectrum measured in reflection ($S_{ii}$) and one measured
in transmission ($S_{ij}$).  The scale of the ordinate is discussed
at Eq.~(\protect\ref{norm}).  
}
\label{fig9}
\end{figure}

We analyze original raw spectra of the hyperbola with a total number
of $\Lambda=3$ antennas. Besides the unfolding, no further preparation
has been performed. On Fig.~\ref{fig9}, the function
$\langle|C(t)|^2\rangle$ is shown for raw spectra of the hyperbola
billiard.  The ordinate is in principle as in Figs.~\ref{fig4},
\ref{fig5} and \ref{fig8}: The function $\langle|C(t)|^2\rangle$ is
given in units of $N\overline{y^2}$ in order to allow immediate
comparison with Fig.~\ref{fig3} and all other similar figures, see
Eq.~(\ref{rawfour}). It is, however, not obvious how to obtain
$N\overline{y^2}$ from a spectrum unless one identifies and analyzes
all the resonances. We proceed as follows.

If one assumes the line shape $L(x)$ to be a Lorentzian with width 
$\Gamma$, then the integral over the square of the spectrum $I(x)$ of 
Eq.~(\ref{intensity}) is
\begin{equation}
       \int_{-\infty}^{+\infty} I^2(x) dx
       = \frac{4\pi}{\Gamma^3} \sum_{\mu=1}^N y_\mu^2 
       \approx 4\pi \frac{N\overline{y^2}}{\Gamma^3} \ .
       \label{norm}
\end{equation}
This quantity is obtained by numerical integration. The width $\Gamma$
is obtained by a fit--procedure from the exponential decay of
$\langle|C(t)|^2\rangle$ for $t>1$ as is illustrated by
Fig.~\ref{fig9}. See also the discussion at the end of
Sec.~\ref{sec3a}. From these two pieces of information,
Eq.~(\ref{norm}) yields $N\overline{y^2}$.

In the upper part of Fig.~\ref{fig9}, the function
$\langle|C(t)|^2\rangle$ is shown for a raw spectrum of the hyperbola
measured in reflection mode. Unfortunately, this function cannot be
interpreted in terms of the correlation hole because it is dominated,
up to $t\approx 2$ by a background which is typical for the reflection
measurements. Since the radio frequency cables we used for the
transmission between source and resonator are interrupted by certain
non-ideal cable connectors, the measured frequency spectrum is
modulated by the signal that is generated through reflections at these
connectors and the resonator; see the oscillations of the spectral
background in Fig.~1 of Ref.~\cite{widths}.  Furthermore, the size of
the resonators and the lengths of the cables are of the same order.
Therefore, the period of this spectral modulation is of the order of
the mean level spacing and the power spectrum between $t=0$ and
$t\approx 2$ is dominated by this artificial background peak.
Note, however, that this does not at all preclude the analysis
of every given resonance, because $\Gamma$ is very small as compared
to the period of the above oscillations.

\begin {figure}
\centerline{\epsfxsize=8.6cm
\epsfbox{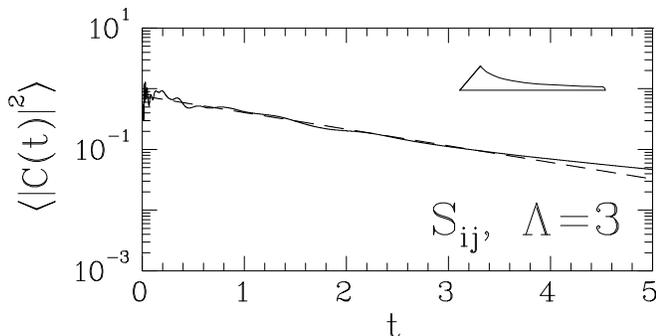}}
\caption{
The function $\langle |C(t)|^2 \rangle$ for 
three superposed transmission spectra of the hyperbola.
The scale of the ordinate is discussed
at Eq.~(\protect\ref{norm}).  
}
\label{fig10}
\end{figure}

The transmission spectra are free from that problem. Nevertheless, the
correlation hole is not visible in the lower part of Fig.~\ref{fig9},
where $\langle|C(t)|^2\rangle$ is given for a raw spectrum of the
hyperbola measurement in transmission mode. At the place of the
correlation hole, one observes fluctuations that will be discussed in
Sec.~\ref{sec4c3} below. They are due to the statistical fluctuations
of the intensities $y_\mu$. Note that the value of $\alpha=1/9$,
expected from Eq.~(\ref{tranalpha}) for this case, leaves little hope
to see the correlation hole. According to Sec.~\ref{sec3d}, the value
of $\alpha$ should improve if several spectra are superimposed. We
have done so for all transmission spectra ($\Lambda=3$) of the
hyperbola. According to Eq.~(\ref{tranalphatot}), one then expects
$\alpha=1/5$. Figure~\ref{fig10} shows that this is not enough. 
As compared to the lower part of Fig.~\ref{fig9}, the fluctuations
are more suppressed, but the hole itself cannot be identified.

\subsubsection{Idealized Raw Spectra} 
\label{sec4c2}

\begin {figure}
\centerline{\epsfxsize=8.6cm
\epsfbox{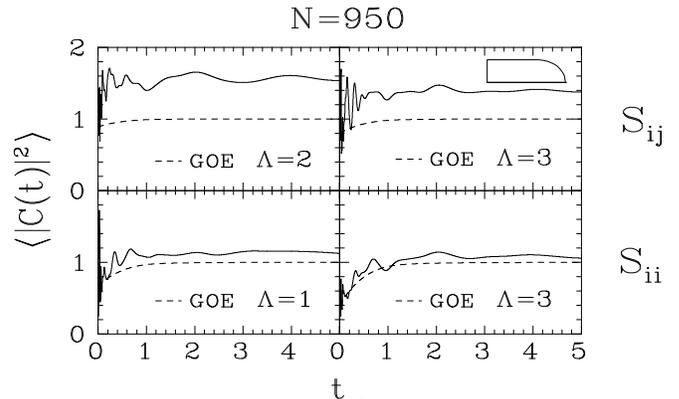}}
\caption{
The function $\langle |C(t)|^2 \rangle$ for an idealized raw 
spectrum of the stadium-billiard, as in Fig.~4. See the detailed 
explanation in the text. The experimental curves are given as 
solid, the theoretical ones as dashed lines.  
}
\label{fig11}
\end{figure}

In order to acquire a deeper understanding of the statistical effects
that are important for the correlation--hole method, we study 
``idealized raw spectra''. They are obtained by providing the
experimentally found stick--spectra with realistic intensities: We use
the experimentally determined spectral properties of the
stadium-billiard with $\Lambda=3$ channels, i.e. the full set of
parameters for the first 950 resonances including their positions
$x_\mu$ as well as three sets of partial widths $\Gamma_{\mu c}$ which
were shown to obey the Porter-Thomas distribution~\cite{widths}. Note
that these partial widths have been separately normalized to unity for
each channel, i.e. $\langle\Gamma_{\mu c}\rangle=1$.  The resulting
functions $\langle |C(t)|^2 \rangle$ for different cases are given in
Fig.~\ref{fig11}. In the upper half of the figure, transmission
spectra $S_{ij}$ for $i \neq j$ are analyzed. In the upper left part,
a single transmission spectrum is used. This means $\Lambda=2$ in
Eq.~(\ref{sechszwanz}) and the theory predicts $\alpha=1/9$.  In the
upper right part, three transmission spectra have been superposed,
i.e.~$\Lambda=3$ and the prediction is $\alpha=1/5$.  In the lower
half of the figure, reflection spectra $S_{ii}$ are analyzed. A single
reflection spectrum is used in the lower left part.  This means
$\Lambda=1$ in Eq.~(\ref{fuenfzwanz}) yielding $\alpha=1/3$.  In the
lower right part, $\Lambda=3$ reflection spectra have been superposed
and Eq.~(\ref{fuenfzwanz}) predicts $\alpha=3/5$.  Note in addition,
that $\langle |C(t)|^2 \rangle$ as well as the function in
Eq.~(\ref{rawfour}) were divided by $N\overline{y^2}$.  Obviously,
there is a strong deviation between the experimental and the
theoretical curve in the case of the single spectra in the left column
of Fig.~\ref{fig11}.  For the superpositions, this deviation is
reduced.

\subsubsection{Statistical Fluctuations of the Squared Intensities
               and their Impact}
\label{sec4c3}

To give a qualitative interpretation of this deviation, we study
the statistical fluctuations of $|C(t)|^2$ in the long-time-limit $t
\to \infty$, where the power spectrum is free of effects due to level
clustering.  We will show that the deviation just observed can be
attributed to, at first sight unexpectedly, large statistical
fluctuations of the squared intensities in the spectra.
From Eqs.~(\ref{intensity})
and (\ref{autocorri}) with $L(x)=\delta(x)$ one obtains
\begin{equation}
     |C(t)|^2 = \sum_{\mu=1}^N y_\mu^2+
        \sum_{{\mu,\nu=1} \atop {\mu \ne \nu}}^N 
           y_\mu y_\nu \exp\big(2\pi i (x_\mu-x_\nu)t\big) \ .
     \label{z1}
\end{equation}
In the limit $t \to \infty$ and due to the full Gaussian smoothing,
the strongly fluctuating second term of this expression is
suppressed. We define this limit as
\begin{equation}
  X = \lim_{t \to \infty} \langle |C(t)|^2 \rangle = 
              \sum_{\mu=1}^N y_\mu^2 \ .
     \label{z2}
\end{equation}
The aforementioned ensemble average in the limit $N\to\infty$
yields the quantity
\begin{equation}
     \overline{X} = \lim_{N \to \infty} X = N\overline{y^2}
     \label{z3}
\end{equation}
which is exactly the expression we used for the normalization in
the previous analysis of the idealized raw spectra.
Now, to estimate the statistical fluctuations of $X$,
we have to calculate its second moment $\overline{X^2}$ 
and the relative standard deviation
\begin{equation}
    \delta_{\rm rel} X = 
     \sqrt{\frac{\overline{X^2}-\overline{X}^2}
           {\overline{X}^2}} \ .
     \label{z4}
\end{equation}
In the case of a single transmission spectrum $S_{ij}$ with $\Lambda=2$,
one has $y_\mu=\Gamma_{\mu a}\Gamma_{\mu b}$ which yields
\begin{equation}
     \delta_{\rm rel} X = \frac{1}{\sqrt{N}} \,
       \sqrt{\frac{\overline{\Gamma^4}^2}{\overline{\Gamma^2}^4}
                                       -1} \ .
     \label{z5}
\end{equation}
This behaves, as expected, like $1/\sqrt{N}$. However, since
the partial widths are Porter-Thomas distributed,
the higher order moments of $\Gamma_\mu$ obey
\begin{equation}
   \overline{\Gamma^k} = (2k-1)!! \, \overline{\Gamma}^k 
     \label{z6}
\end{equation}
implying that the statistical fluctuations strongly increase with the
order of the statistical moment due to the factor $(2k-1)!!$. In the
present case of $N=950$ resonances one obtains $\delta_{\rm rel} X
\approx 0.38$ which is in good agreement with the experimental curve
given in Fig.~\ref{fig11}.  For an increasing number of channels
$\Lambda$, this relative variance shrinks, since the superposition
reduces the fluctuations in the intensities $y_\mu$. For vanishing
fluctuations in the intensities, one has $\delta_{\rm rel}X=0$. Since
the parameter $\alpha$ is generated through the statistical moments of
the $y_\mu$, both, the normalization as well as the correlation hole
itself, approach the theoretical prediction only as the number of
included open channels is increased.

Remarkably, a superposition of reflection spectra even for the small
number of only $\Lambda=3$ open channels allows, in principle, a
re--observation of the correlation hole since we have $\alpha=3/5$. For
the superimposed transmission spectra with $\alpha=1/5$, the hole is
still weak.  This explains why the hole could not be observed in the
superposition of the original transmission spectra of the hyperbola in
Fig.~\ref{fig10}.  Unfortunately, the favorable behavior of the
reflection measurements cannot be exploited because of the
experimental artifact discussed in Sec.~\ref{sec4c1}.

\begin {figure}
\centerline{\epsfxsize=8.6cm
\epsfbox{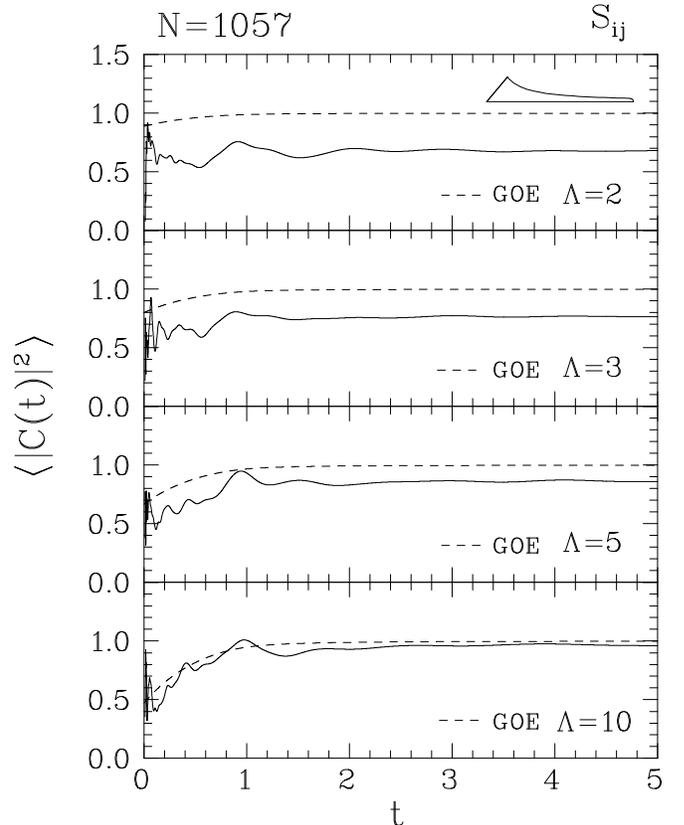}}
\caption{
Dependence of the visibility of the correlation hole on the number
$\Lambda$ of open channels for $\Lambda=2,3,5$ and $10$.
The scale of the ordinate is as in Fig.~4.
}
\label{fig12}
\end{figure}

In order to demonstrate the influence of the number $\Lambda$ of open
channels on the correlation hole and on the normalization, we have
finally calculated several superpositions of synthetic spectra for the
hyperbola. We used the experimental spectrum of the positions $x_\mu$
and up to $\Lambda=10$ numerically simulated spectra of Porter-Thomas
distributed partial widths $\Gamma_{\mu}$ with $\langle\Gamma_{\mu
  c}\rangle=1$. Thus, in contrast to the idealized raw spectra of the
Sec.~\ref{sec4c2} where the amplitudes were taken from the
measurement, the synthetic spectra contain numerically generated
amplitudes. Figure~\ref{fig12} displays the results for
$\Lambda=2,3,5$ and 10. Since we treated the spectra as measured in
transmission mode, we have, according to Eq.~(\ref{sechszwanz}),
ratios of the moments given by $\alpha=1/9,1/5,1/3$ and $9/17$,
respectively.  Obviously, the systematic increase in $\Lambda$ leads
to a continuing reduction of the statistical fluctuations of $y_\mu$,
described by $\delta_{\rm rel}X$. For $\Lambda=10$ the agreement
between the experimental and the theoretical curve is quite good.
This demonstrates, for transmission and reflection measurements, how
the correlation hole can be observed by superimposing independent
spectra which, individually, do not show a significant correlation
hole.

\section{Conclusion}
\label{sec5}

In the present work on the correlation--hole method, we have evaluated
the decay function $\langle |C(t)|^2 \rangle$ of stick--spectra as
well as that of original and idealized raw spectra for various two--
and three--dimensional billiard systems.

The stick--spectra lead to the correlation hole as expected from
Random Matrix Theory. The non--generic features implied by the
presence of bouncing ball orbits in certain billiards did not affect
the correlation hole in any observable way.

The fluctuations of the resonance intensities that are present in the
raw spectra have a strong impact on the results. They decrease the
visibility of the correlation hole. At the same time, they introduce
fluctuations such that --- even with approximately 1000 resonances in
the spectrum --- the decay function $\langle |C(t)|^2 \rangle$ may
fall quite far from its expected shape. This again precludes the
observation of the correlation hole. It is, however, restored if
sufficiently many spectra with statistically independent intensities
are superimposed.

\section*{Acknowledgments}

We would like to thank H. Lengeler and the CERN workshops for the
excellent fabrication of the niobium resonators.  We are grateful to
J. Nyg\aa rd for helpful advice regarding the smoothing procedure of
the Fourier transforms. We thank T. Gorin and I. Rotter for a very
useful conversation. We also benefited from fruitful discussions with
C. Dembowski and with C. Ianes Barbosa who was supported from a grant
of the Deutsche Akademische Austauschdienst during her stay in
Darmstadt.  TG acknowledges financial support from a
Habilitanden--Stipendium of the Deutsche Forschungsgemeinschaft. This
work has been supported by the Sonderforschungsbereich 185
``Nichtlineare Dynamik'' of the Deutsche Forschungsgemeinschaft and in
part by the Bundesministerium f\"ur Bildung und Forschung under
contract number 06DA665I.

\end{multicols}


\begin{references}

\bibitem{SS9092}     H.-J. St\"ockmann and J. Stein, 
                     Phys. Rev. Lett. {\bf 64}, 2215 (1990);
                     J. Stein and H.-J. St\"ockmann, 
                     Phys. Rev. Lett. {\bf 68}, 2867 (1992).
\bibitem{DSF90}      E. Doron, U. Smilansky, and A. Frenkel,
                     Phys. Rev. Lett. {\bf 65}, 3072 (1990).
\bibitem{Sridhar91}  S. Sridhar, 
                     Phys. Rev. Lett. {\bf 67}, 785 (1991).
\bibitem{Grapa}      H.-D. Gr\"af, H.L. Harney, H. Lengeler, 
                     C.H. Lewenkopf, C. Rangacharyulu, A. Richter, 
                     P. Schardt, and H.A. Weidenm\"uller,
                     Phys. Rev. Lett. {\bf 69}, 1296 (1992).
\bibitem{Porter}     C.E. Porter, 
                     {\it Statistical Theories of Spectra: 
                     Fluctuations}, (Academic Press, 1965).
\bibitem{Mehta}      M.L. Mehta, 
                     {\it Random Matrices}, 2nd ed.,
                     (Academic Press, San Diego, 1991).
\bibitem{BG84}       O. Bohigas and M.-J. Giannoni,
                     in {\it Chaotic Motion and Random Matrix Theory},
                     Lecture Notes in Physics Vol.~209 
                     (Springer, Berlin, 1984), p.1.
\bibitem{Bohigas}    O. Bohigas, 
                     in {\it Chaos and Quantum Physics}, 
                     M.-J. Giannoni, A. Voros, and J. Zinn-Justin, eds.
                     (Elsevier, Amsterdam, 1991), p. 89.
\bibitem{Berry}      M. Berry, 
                     in {\it Chaos and Quantum Physics}, 
                     M.-J. Giannoni, A. Voros, and J. Zinn-Justin, eds.
                     (Elsevier, Amsterdam, 1991), p. 253.
\bibitem{Haake}      F. Haake,
                     {\it Quantum Signatures of Chaos}
                     (Springer, Berlin, 1991).     
\bibitem{Leviandier} L. Leviandier, M. Lombardi, R. Jost, and J.P. Pique,
                     Phys. Rev. Lett. {\bf 56}, 2449 (1986).
\bibitem{Delon}      A. Delon, R. Jost, and M. Lombardi, 
                     J. Chem. Phys. {\bf 95}, 5701 (1991).
\bibitem{BohigasGSE} M. Lombardi, O. Bohigas, and T.H. Seligman, 
                     Phys. Lett. B{\bf 324}, 263 (1994).
\bibitem{Guhr1}      T. Guhr and H.A. Weidenm\"uller, 
                     Chem. Phys. {\bf 146}, 21 (1990).
\bibitem{Guhr2}      U. Hartmann, H.A. Weidenm\"uller, and T. Guhr,
                     Chem. Phys. {\bf 150}, 311 (1991).
\bibitem{LPLBS}      M. Lombardi, J.P. Pique, P. Labastie,
                     M. Broyer and T. Seligman,
                     Comments. At. Mol. Phys. {\bf 25}, 345 (1991).
\bibitem{LoSe}       M. Lombardi and T.H. Seligman,
                     Phys. Rev. {\bf A47}, 3571 (1993).
\bibitem{Alhassid}   Y. Alhassid and N. Whelan,  
                     Phys. Rev. Lett. {\bf 70}, 572 (1993).
\bibitem{Sridhar}    A. Kudrolli, S. Sridhar, A. Pandey, and R. Ramaswamy,
                     Phys. Rev. E{\bf 49}, R11, (1994).
\bibitem{Hyperbel}   H. Alt, H.-D. Gr\"af, H.L. Harney, R. Hofferbert,
                     H. Lengeler, C. Rangacharyulu, A. Richter, 
                     and P. Schardt,
                     Phys. Rev. E{\bf 50}, 1 (1994).
\bibitem{widths}     H. Alt, H.-D. Gr\"af, H.L. Harney, R. Hofferbert, 
                     H. Lengeler, A. Richter, P. Schardt,
                     and H.A. Weidenm\"uller,
                     Phys. Rev. Lett. {\bf 74}, 62 (1995).
\bibitem{Beschl}     J. Auerhammer, H. Genz, H.-D. Gr\"af, R. Hahn,
                     P. Hoffmann-Stascheck, C. L\"uttge, U. Nething, 
                     K. R\"uhl, A. Richter, T. Rietdorf, P. Schardt, 
                     E. Spamer, F. Thomas, O. Titze, J. T\"opper,
                     and H. Weise,
                     Nucl. Phys. A{\bf 553}, 841c (1993).
\bibitem{Primack}    H. Primack and U. Smilansky, 
                     Phys. Rev Lett. {\bf 74}, 4831 (1995).
\bibitem{3DSinai}    H. Alt et al., to be published.
\bibitem{Weaver}     R.L. Weaver, 
                     J. Acoust. Soc. Am. {\bf 85}, 1005 (1989).
\bibitem{ACME}       C. Ellegaard, T. Guhr, K. Lindemann, H.Q. Lorensen,
                     J. Nyg\aa rd, and M. Oxborrow,
                     Phys. Rev. Lett. {\bf 75}, 1546 (1995).
\bibitem{quartz}     C. Ellegaard, T. Guhr, K. Lindemann, 
                     J. Nyg\aa rd and M. Oxborrow, 
                     Phys. Rev. Lett. {\bf 77}, 4918 (1996).
\bibitem{Brentano}   H. Alt, P. von Brentano, H.-D. Gr\"af,
                     R.-D. Herzberg, M. Philipp, A. Richter,
                     and P. Schardt,
                     Nucl. Phys. A {\bf 560}, 293 (1993).
\bibitem{Brentanoneu}H. Alt, P. von Brentano, H.-D. Gr\"af, R. Hofferbert,
                     M. Philipp, H. Rehfeld, A. Richter, and P. Schardt,
                     Phys. Lett. B {\bf 366}, 7 (1996).
\bibitem{Weyl1}      H. Weyl, 
                     Journal f\"ur die reine und angewandte Mathematik,
                     {\bf 141}, 1 (1912); {\it ibid.} p.~163.
\bibitem{Weyl2}      H. Weyl, 
                     Journal f\"ur die reine und angewandte Mathematik,  
                     {\bf 143}, 177 (1913).
\bibitem{BalHilf}    H.P. Baltes and E.R. Hilf, 
                     {\it Spectra of Finite Systems}
                     (Bibliographisches Institut, Mannheim, 1975).
\bibitem{MW}         C. Mahaux and H.A. Weidenm\"uller,
                     {\it Shell-Model Approach to Nuclear Reactions}
                     (North-Holland, Amsterdam, 1969).
\bibitem{Gutzwiller} M.C. Gutzwiller, 
                     {\it Chaos in Classical and Quantum
                     Mechanics} (Springer, New York, 1990).
\bibitem{DysonMehta} M.L. Mehta and F.J. Dyson, 
                     J. Math. Phys. {\bf 4}, 713 (1963).
\bibitem{Sieber}     M. Sieber, 
                     dissertation (Universit\"at Hamburg, 1990).
\bibitem{Hesse}      R. Aurich, T. Hesse, and F. Steiner, 
                     Phys. Rev. Lett. {\bf 74}, 4408 (1995).
\bibitem{BerryGOE}   O. Bohigas, M.J. Giannoni, and C. Schmit, 
                     Phys. Rev. Lett. {\bf 52}, 1 (1984).
\bibitem{Bunimovich} L.A. Bunimovich, 
                     Sov. Phys. JETP {\bf 62}, 842 (1985).
\bibitem{Smilansky}  M. Sieber, U. Smilansky, S.C. Creagh,
                     and R.G. Littlejohn, 
                     J. Phys. A {\bf 26}, 6217 (1993).

\end{references}
\end{document}